\documentstyle[aas2pp4]{article}

\begin{document}

\title{Ultraviolet Signatures of Tidal Interaction \\ 
in the Giant Spiral Galaxy, 
M101}
\author{William H. Waller\altaffilmark{1}, 
Ralph C. Bohlin\altaffilmark{2}, 
Robert H. Cornett\altaffilmark{1}, 
Michael N. Fanelli\altaffilmark{1},
Wendy L. Freedman\altaffilmark{3}, 
Jesse K. Hill\altaffilmark{1}, 
Barry F. Madore\altaffilmark{4}, 
Susan G. Neff\altaffilmark{5}, 
Joel D. Offenberg\altaffilmark{1}, 
Robert W. O'Connell\altaffilmark{6}, 
Morton S. Roberts\altaffilmark{7}, 
Andrew M. Smith\altaffilmark{5}, 
and Theodore P. Stecher\altaffilmark{8}}

\altaffiltext{1}{Hughes STX Corporation, NASA Goddard Space Flight Center,
Laboratory for Astronomy and Solar Physics, Code 681, Greenbelt, MD  20771}

\altaffiltext{2}{Space Telescope Science Institute, 3700 San Martin Drive,
Baltimore, MD  21218}

\altaffiltext{3}{Carnegie Observatories, 813 Santa Barbara St., Pasadena, CA
91101-1292}

\altaffiltext{4}{Infrared Processing and Analysis Center, Caltech, M/S
100-22, 770 So. Wilson Ave., Pasadena, CA  91125}

\altaffiltext{5}{NASA Goddard Space Flight Center, Laboratory for Astronomy
and Solar Physics, Code 681, Greenbelt, MD  20771}

\altaffiltext{6}{University of Virginia, Department of Astronomy, P. O. Box
3818, Charlottesville, VA  22903}

\altaffiltext{7}{National Radio Astronomy Observatory, 520 Edgemont Rd.,
Charlottesville, VA  22903--2475}

\altaffiltext{8}{NASA Goddard Space Flight Center, Laboratory for Astronomy
and Solar Physics, Code 680, Greenbelt, MD  20771}

\begin{abstract}

We present new evidence for tidal interactions having occurred in the disk of
M101 in the last 10$^8$ -- 10$^9$ years.  Recent imaging of the far-ultraviolet
emission from M101 by the Shuttle-borne Ultraviolet Imaging Telescope (UIT)
reveals with unprecedented clarity 
a disk-wide pattern of multiple linear arm segments (``crooked arms'').
The deep FUV image also shows a faint outer spiral arm with a 
(``curly tail'') feature  
that appears to loop around the supergiant HII region NGC 5471 --- 
linking this outlying
starburst with the rest of the galaxy.  These FUV-bright features most likely
trace
hot O \& B-type stars along with scattered light from associated nebular dust.
Counterparts of the outermost ``crooked arms''
are evident in maps at visible wavelengths and in the 21-cm line of HI.  
The inner-disk FUV arms are most closely associated with H$\alpha$ knots and 
the outer (downstream) sides of CO arms.
Comparisons of the ``crooked arm'' and ``curly tail''
morphologies
with dynamical simulations yield the greatest similitude, when the
non-axisymmetric forcing comes from a combination of {\it external interactions}
with one or more companion galaxies and {\it internal perturbations}
from
massive objects orbiting
within the
disk.  We speculate that NGC 5471 represents one of these ``massive
disturbers'' within the disk,
whose formation followed from a tidal interaction between M101 and a smaller
galaxy.

\end{abstract}


\section{Introduction}
Recent HST measurements of Cepheid variables in M101 
(\markcite{kelson}Kelson et al. 1996)
have shown that this Sc(s)I
galaxy is $\sim$50\%
more distant than was previously inferred from groundbased measurements of
the galaxy's brightest supergiant stars 
(\markcite{ha}Humphreys \& Aaronson 1987).
At the new distance of 7.4 $\pm$
0.60 Mpc, M101 has a Holmberg radius of
30.1 kpc (\markcite{Holmberg}Holmberg 1958), while
its outermost giant HII regions lie more than 25 kpc from its nucleus.  A
galaxy of such enormous extent (70\% larger than the giant ScI galaxy M100) 
presents an unusually large cross-section for
interactions with its galaxian neighbors.  The presence of the companion dwarf
galaxies NGC 5477 and NGC 5474 with projected radii of 47 kpc and 99
kpc, respectively, lends further credence to the idea of
tidal interactions having recently modified the disk of M101
(cf. \markcite{vdh}van der Hulst \& Sancisi 1988;
\markcite{ksw}Kenney et al. 1991; \markcite{combes}Combes 1991).

Evidence for interaction events during the last Gyr include (1) the lopsided
distribution of HI gas (towards the northeast) with respect to the galaxy's
dynamical nucleus (\markcite{bosma}Bosma et al. 1981), (2) deviations from 
circular rotation in the outer galaxy (\markcite{bosma}Bosma et al. 1981; 
\markcite{kamphuis}Kamphuis 1993) (3) massive
(10$^7$ -- 10$^8$ M$_{\odot}$)
high-velocity clouds of HI gas to the east of the nucleus along with contiguous
``holes'' of comparable size (\markcite{vdh}van der Hulst \&
Sancisi 1988),
(4) a non-axisymmetric distribution of O/H abundances
(\markcite{kl}Kennicutt
\& Liu 1996), and (5) outer spiral arms with nearly linear
morphologies
to
the east and with bifurcating non-logarithmic curvatures to the west.  
The unusual arm
morphology to the east has been described in a variety of terms, including
``crooked arms'' (\markcite{gb}Garcia-Burillo et al.1991),
``kinks'' (\markcite{eem}Elmegreen et al. 1992),
``spurs'' (\markcite{byrd}Byrd 1983; \markcite{bsm}Byrd et al. 1984),
``squared-off arm sections'' and ``intermediate-scale structure''
(\markcite{r92}Roberts 1992; \markcite{r93}Roberts 1993).

Similar non-logarithmic spiral structure has been
noted in the eastern-most arm of M51, whose obvious interaction with NGC 5195
has prompted the development of dynamical models, some of which successfully
predict the generation of distorted spiral structure 
(\markcite{gb}Garcia-Burillo et al. 1991; 
\markcite{ekt}Elmegreen et al. 1993;
\markcite{bs}Byrd \& Salo 1995).  Other numerical
simulations
involving massive objects orbiting {\it within} the disks of spiral galaxies
yield transient crooked arms, kinks, and spurs 
(\markcite{bsm}Byrd et al. 1984; 
\markcite{r92}Roberts 1992).

In this paper, we present new evidence for dynamical interactions having
occurred in the disk of M101 during the last 10$^8$ -- 10$^9$ years.
Recent imaging of the
far-ultraviolet emission from M101
by the Shuttle-borne {\it Ultraviolet Imaging Telescope}
reveals a remarkable pattern of 
multiple linear arm segments (``crooked arms'')
throughout the disk and a faint outer spiral
arm with a (``curly tail'') feature 
that appear to be of tidal origin.
These FUV-prominent features most likely
trace hot O \& B-type stars along with scattered light from associated
concentrations
of dust.  The morphological results presented herein build
upon those obtained at lower resolution 
in a previous rocket flight, where the larger-scale UV
emission is described (\markcite{s82}Stecher et al. 1982).

Section 2 briefly describes the {\it Astro-2/UIT} observations and reductions.
The FUV-emitting features
are presented in Section 3.  The interpretation of these features as
physically significant structures is evaluated in Section 4, and the dynamical
implications are discussed in Section 5.  Subsequent papers will present more
detailed analyses of the FUV emission features
in M101 (\markcite{w96}Waller et al. 1996) and of
the dynamical processes which these features trace 
(\markcite{cow}Combes et al. 1996).
 
\section{Observations and Reductions}
The Ultraviolet Imaging Telescope imaged M101 in the far-ultraviolet
($\lambda_{\circ}$ = 1521 \AA, $\Delta\lambda$ = 354 \AA) on 11, March 1995
as part of the 16-day Spacelab/{\it Astro-2} mission aboard the Space Shuttle
Endeavour.  The 1310-sec exposure was
obtained with a dual-stage image intensifier with CSI photocathodes
and was recorded on carbon-backed
IIaO Kodak film.  After processing of the film and scanning of the emulsion,
the
resulting digitized ``density image'' was fog-subtracted, flat-fielded,
linearized to ``exposure
units,'' and calibrated to flux units
using IUE observations of standard stars (cf. \markcite{s92}Stecher
et al. 1992; \markcite{w95}Waller et al. 1995).
Correction for image
distortion produced by the magnetically focused image intensifiers
was carried out according to the procedures described by 
\markcite{greason}Greason et
al. (1995).  The resulting corrections amounted to a few arcsec in the field
center increasing to 10--20 arcsec near the edge of the 40 arcmin field of 
view.
The astrometry of the distortion-corrected image is good to $\sim$3
arcsec, and the spatial resolution is of
similar magnitude.
 
\section{Morphology of FUV Emission}
{\bf Figure 1} (Plate xxxx) shows the distortion-corrected
FUV image of M101.  The
emission morphology is dominated by spiral arms and star-forming knots.  Most
of the FUV-bright knots coincide with H$\alpha$-bright giant HII regions (cf.
\markcite{w90}Waller 1990; \markcite{sdh}Scowen et al. 1992).
However, the FUV knots are connected by brighter lanes of emission.
Some of this FUV emission can be attributed to scattering of starlight
by dust along the
spiral arms (\markcite{s82}Stecher et al. 1982; 
\markcite{hill}Hill et al. 1995), although the balance of
scattered vs. direct FUV emission remains uncertain (see next Section).

The structure of the FUV arms deviates markedly from that of logarithmic
spirals.  Instead, multiple ``crooked arms''
 are
evident throughout the disk of M101.  The ``crooked arm'' morphology consists
of linear arm
segments that intersect one another --- often at angles of
$\sim$120$^{\circ}$.  These patterns are identified in {\bf Figure 2a} (Plate
xxx),
where the FUV
image of M101 has been spatially filtered to highlight the intermediate-scale 
structure.  

The
face-on locations, pitch angles, lengths, surface brightesses, and contrasts
of the 12 arm
segments are listed in {\bf Table 1}, where an 
inclination of 18$^{\circ}$ and major axis position angle of 39$^{\circ}$ 
have been assumed for the rectification
(\markcite{bga}Bosma et al. 1981).
A wide range of mean position angles and radii are represented in the list, 
further highlighting the ubiquitous nature of the linear arm segments.
The large and varying pitch angles
indicate arm morphologies that strongly differ from logarithmic spirals.  The 
background-subtracted FUV 
surface brightnesses of the arm segments 
average to $(5.6 \pm 3.2) \times 10^{-18}$ erg cm$^{-2}$ 
s$^{-1}$ arcsec$^{-2}$ \AA$^{-1}$, which corresponds to 
$2.9 \times 10^{37}$ erg s$^{-1}$ \AA$^{-1}$ 
kpc$^{-2}$ in the disk of the galaxy, 
or the equivalent of 300 Orion nebulae per square kpc 
(\markcite{bohlin}Bohlin et al. 1982).  
The contrast (ratio) of arm-interarm median intensities ranges from 1.3 $\pm$
0.2 for the
innermost arm segment (\#5) to $\ge$ 5.0 $\pm$ 0.2 for the prominent arms in
the east --- where the uncertainties
refer to the rms dispersion of contrasts derived
from an arbitrary sequence of strip scans
across the galaxy.

In addition to the crooked arms, a faint arm evident to the south is seen to
extend all the way to the outermost giant HII region, NGC 5471,
where it appears to loop around and ultimately connect with this
kpc-size starbursting region (see {\bf Figure 2b} [Plate xxx]).
Previous UV imaging (\markcite{s82}Stecher et al. 1982; \markcite{b90}
Blecha et al. 1990) detected
the faint southern arm
(see features B\&C in \markcite{s82}Stecher et al. 1982) but were 
of insufficient resolution ($\sim$20$''$) to delineate the
``curly tail'' structure which appears to connect with NGC 5471.

\section{Physical Interpretation of FUV Features}

Through FUV imaging of M101 at $\sim$3-arcsec resolution, 
an anomolous pattern of non-logarithmic arms 
can now be traced with unprecedented clarity.  The physical significance 
of this FUV morphology, however, is less obvious.  
We note that 
the physical interpretation of other 
spatial patterns in the sky has sometimes led
to
spurious conclusions.  Well-known examples include the interpretation of dark
patterns on Mars as evidence for life-sustaining canals (cf. 
\markcite{hoyt}Hoyt 1976)
and the more
recent controversy over stellar ring configurations seen on
the Palomar Sky Survey images (cf. \markcite{vb}Vidal \& Bern 1973).
Because of the human eye-brain's propensity to ``connect the dots'' and see
patterns where none actually exist,
the ``crooked arm'' and ``curly tail'' features noted in the
previous section are subject to similar mis-interpretation.
Fortunately, these interpretive difficulties can be overcome by comparing 
the FUV features in M101
with 
stellar and interstellar tracers at other wavelengths.

\subsection{Arms and Knots at other Wavelengths}
``Kinks'' and ``spurs'' in the spiral arms of M101 have been previously noted
from optical images (\markcite{bga}Bosma et al. 1981; 
\markcite{eem}Elmegreen et al. 1992).
These descriptions invariably referred to the {\it outermost arms} to the east
and northeast (arm segments \#10,11,12 in {\bf Figure 2a} and {\bf Table 1}),
where confusion with the underlying disk is minimal.  At smaller
galactocentric radii, emission from the underlying disk and absorption by lanes
of dust complicate the observed morphology at blue--red wavelengths.

Comparison of the FUV image
with the blue photograph of M101
printed in the
Atlas of Galaxies (\markcite{sb}Sandage \& Bedke 1988) shows that the
FUV linear arm
segments
in the inner galaxy
{\it do not} coincide with the spiral lanes of dust that are seen in silhouette 
against the brighter disk.
Better correspondence is evident between the FUV-bright
knots making up the arm segments and the young star clusters revealed in the
blue print.
Continuum-subtracted H$\alpha$ emission images show
the star-forming knots in the inner galaxy
with equivalent contrast to that observed in the FUV image
(\markcite{w90}Waller 1990; \markcite{sdh}Scowen et al. 1992).
However, the diffuse inter-knot emission along the arms is less
evident at H$\alpha$ than in the FUV, making the ``crooked-arm'' morphology
less compelling.  The connecting emission seen in the FUV but not at H$\alpha$
is probably tracing
{\it non-ionizing} populations of B-type stars along with associated reflection
nebulosity.

The FUV-bright knots along the arms have similar sizes at
H$\alpha$ and in the FUV.  By contrast, the sizes of the knots in the
red-continuum are smaller by factors of
$\sim$2 (\markcite{w90}Waller 1990).  
Since the red-continuum is probably tracing direct
starlight from the underlying OB clusters, while the H$\alpha$ emission arises 
from nebular ionized gas, the larger FUV and H$\alpha$ sizes
indicate that the FUV emission from the
knots is at least partly nebular
(e.g. reflection nebulosity associated with the
HII regions).  For HII regions and reflection nebulae in the Milky Way, 
scattering efficiencies by nebular dust
are $\sim$3 times greater in the FUV than at visible wavelengths 
(\markcite{witt}Witt et al. 1992), thus explaining the larger FUV sizes.  
By comparing the
peak and total emission from 3 bright knots in the R and FUV images (at 
matching
resolution), we estimate that at least
30\% of the FUV emission is scattered by nebular dust.  UV images of the
Orion nebula
(\markcite{bohlin}Bohlin et al.
1982) and the giant HII regions in M81 (\markcite{hill}Hill et al. 1995)
show evidence for similar amounts of scattering.

Comparison of the UIT's FUV image with maps of 21cm HI emission at
45'' resolution (\markcite{bga}Bosma et al. 1981), 80''
resolution (\markcite{vdh}van der Hulst \& Sancisi 1988), and $13'' \times 16''$
resolution (\markcite{ksh91}Kamphuis et al. 1991; 
\markcite{kamphuis}Kamphuis 1993)
shows that the outermost straight-arm segments
to the east and northeast (\#9,10,11,12) coincide with linear
ridges of atomic gas.
The inner-disk HI distribution is dominated by clumps and holes with little 
resemblance to the FUV arms.

Complementary information on the morphology
of gas in the inner $4 \times 4$ arcmin (8.6 $\times$ 8.6 kpc)
is obtained from a
recent mapping of 2.6mm CO emission
at $9'' \times 7''$ resolution 
(\markcite{k96}Kenney et al. 1996).  {\bf Figure 3} (Plate
xxx) shows a contour
map of the CO emission superposed on a greyscale representation of the FUV
emission.
Besides resolving the central gaseous bar into discrete features,
the CO map reveals spiral structure beyond the bar, some of which coincides
with prominent FUV emission.  {\it These well-correlated FUV--CO
emission features are
present despite likely concentrations of associated dust} --- further
confirming
the utility of FUV emission as a tracer of high-mass star formation in
disk galaxies of low inclination (\markcite{chen}Chen et al. 1992; 
\markcite{hill}Hill et al. 1995).

The most extensive
FUV--CO correlations are evident
in an arc of shared emission 1--1.5 arcmin to the W--SW of the nucleus, in
linear arm segment \#5 which adjoins the SW end of this arc,
and in a large complex
of knots 1-arcmin to the E--SE.  A significant fraction of the correlated
FUV emission is located on the outer (downstream) sides of the CO arms, thus
suggesting
{\it a spatio-temporal sequence of molecular-cloud aggregation, massive
 star formation, and cloud disruption} that can be modeled in terms of 
density-wave theory.  
Similar spatio-temporal relations between the CO and H$\alpha$ arms 
are discussed by Kenney et al. (1996).

Along the W-SW arc (which shows the most coherent 
sequencing), the azimuthal displacements between the FUV and CO emission amount 
to 0.35 $\pm$ 0.1 kpc --- which when divided by the expected time lag between 
cluster formation and peak FUV brightness ($\sim$3 Myr [Hill et al. 1995]) --- 
yields migration velocities through the spiral density wave of 115 $\pm$ 30 
km/s.  Differencing this velocity with that of the orbiting gas at these radii 
(Bosma et al. 1981; Kenney et al. 1991) results in a pattern speed for the 
density-wave front of 19 $\pm$ 5 km s$^{-1}$ kpc$^{-1}$ with a co-rotation 
radius of 12 $\pm$ 3 kpc (5.5 $\pm$ 1.5 arcmin) --- remarkably close to the 
co-rotation radius determined by Elmegreen et al. (1992) from their multi-mode 
analysis.  The fact that coherent sequencing of the CO and FUV arms 
is not universal to the inner galaxy indicates that other dynamical processes 
are probably competing with the larger-scale density waves.

\subsection{Faint Southern Arm and NGC 5471}

The faint FUV southern arm and the
supergiant HII region (NGC 5471) at the arm's NE tip are
well correlated with other tracers of stellar and
interstellar matter.  The
blue photograph of M101 in the Atlas of Galaxies 
(\markcite{sb}Sandage \& Bedke 1988)
shows the arm to contain many
blue star clusters at the limit of detection.
The supergiant HII region,
NGC 5471,
is beyond the field of the Atlas
print, but appears in the deep blue Schmidt plate
obtained by \markcite{st}Sandage \& Tammann (1974) 
along with diffuse emission to the NW
and $\sim$15 discrete sources to the E and NE (2 of which appear to be
background
galaxies).

The compilation of H$\alpha$
sources by \markcite{hgg}Hodge et al. (1990) 
indicates $\sim$25 sources
associated with the faint southern arm.  Near NGC 5471, their H$\alpha$ image
reveals counterparts to the 4 FUV knots that delineate the tightly
curled part of the
``curly tail'' structure (T1, T2, T3, \& T4 in {\bf Fig. 2a}).
 Beginning 1.8 arcmin to the east of NGC 5471 and looping northward and
westward, the corresponding H$\alpha$ sources are catalogued as \#1258, 1252,
1248, and 1231.  The supergiant HII region, NGC 5471, 
is the most luminous HII region in M101.  
Spectroscopic studies indicate that the underlying stellar population is very 
young ($\leq$3 Myr) resulting in a high H$\alpha$ equivalent width, some 
evidence for an embedded supernova remnant, but little 
evidence for Wolf-Rayet or red supergiant stars (\markcite{s85}Skillman 1985; 
\markcite{rb94}Rosa \& 
Benvenuti 1994).  That NGC 5471 is just now undergoing a massive starburst 
appears to be fortuitous. 

In HI emission, the faint southern arm shows up as a
significant distortion in the HI radial velocity field, deviating from circular
rotation by 10--25 km/sec along the line of sight 
(\markcite{bga}Bosma et al. 
1981).
The $13'' \times 16''$ HI mapping by 
\markcite{ksh}Kamphuis et al. (1991) reveals
an extended ``hole'' of low HI column density
interior to the faint southern arm.  The ``hole'' terminates near NGC 5471,
where
an expansive ($\sim$1 arcmin $\approx$ 2 kpc) region of HI emission is
concentrated.
In the following section, we contend that the velocity
distortions and HI deficits associated with the faint southern arm
are both caused by tidal interaction with
one or more interloping galaxy companions.

\section{Dynamical Implications}

The anomolous morphological and kinematic features in the disk of M101
pose a serious challenge to
dynamicists concerned with the evolution of giant spiral galaxies.  Can a
single model explain the ``crooked arm'' and ``curly tail'' morphologies
seen in the FUV, along
with the lopsided HI distribution, extended HI holes, and contiguous
high-velocity HI features?  
Anticipating numerical simulations specific to the M101 system 
(\markcite{cow}Combes et al. 1996), 
we have examined existing simulations of M101 (\markcite{combes}Combes 1991) 
and of ``generic'' 
disk galaxies to see which ones best
replicate the observed features in M101.

\subsection{External Interactions}

We find that tidal interactions with one or more companion galaxies are able to
produce lopsided
distributions in the gas and stars, large-scale spiral structure with some
cusp-like corners in the gas, and high-velocity features
with contiguous ``holes'' in the gas (cf. \markcite{combes}Combes 1991; 
\markcite{gb}Garcia-Burillo et al. 1991;
\markcite{howard}Howard et al. 1993; \markcite{ekt}Elmegreen et al.
1993; \markcite{bs}Byrd \& Salo 1995).
Simulations with direct encounters relative to the
direction of galactic rotation --- along with large gas fractions and
strong self-gravity within the disks --- yield the clearest spiral structure
(Howard et al. 1993).  Characteristic interaction time scales are of order
$10^8$--10$^9$ years.  For M101,
prime candidates for interaction within this timescale are the dwarf galaxies
NGC 5477 and NGC 5474, whose masses and projected radii are 0.5 $\times$ 10$^9$
M$_{\odot}$ (47 kpc) and 19 $\times$ 10$^9$ M$_{\odot}$ (99 kpc),
respectively (\markcite{allen}Allen et al. 1978; \markcite{combes}Combes 1991).
Such interactions can induce the formation of massive
condensations at the ends of the tidal tails 
(\markcite{ekt}Elmegreen et al. 
1993).  According to this scenario, the kinematically distinct FUV
arm to the south
may represent a tidally induced tail, and the starburst NGC 5471 at the
end of the faint arm may have resulted from a gravitational instability in the
tidal tail.

Further insight to the relevant interaction 
timescales can be gleaned from the geometric 
and kinematic configuration of the tidally induced features.  
In a theoretical study of 
tidal perturbations, gravitational amplification, and galaxy spiral arms, Byrd 
(1995) derives the time since the perturber's closest approach to be 

$$t = R \ (cot \ p) \ / \ v_{rot}$$

\noindent where t is the time since closest approach, p is the pitch angle, and 
v$_{rot}$ is the rotation velocity (see also \markcite{bt}Binney \& Tremaine 
1987).  Using this simple 
formulation, we obtain for the 
faint southern arm a mean interaction 
timescale of $(4.4 \pm 3.3) \times 10^8$ yrs along 
its 19--28 kpc radius.  If we adopt a 27$^{\circ}$ inclination 
(\markcite{Kamphuis}Kamphuis 
1993) instead of the assumed 18$^{\circ}$ (\markcite{Bosma}Bosma 1981), 
the timescale would 
increase to $(6.5 \pm 4.8) \times 10^8$ yrs.  Numerical simulations of 
the interaction between NGC 5474 and M101 
yield very similar timescales (\markcite{cow}Combes et al. 1996).

\subsection{Internal Perturbations}

Simulations involving {\it external interactions} with interloping galaxies
do not usually generate ``crooked arms'' with {\it linear} arm segments
(cf. \markcite{howard}Howard et al 1993).
Multi-mode spiral density waves seem to do a better job of creating breaks and 
kinks in
the spiral structure,
but the arm segments tend to be curved 
(\markcite{eem}Elmegreen et al. 
1992).
Such superpositions of two and three-arm
density-wave modes are more successful at
explaining the broken but
symmetric spiral-arm structure evident in the ``grand design'' spiral
NGC 628
(\markcite{chen}Chen et al. 1992; \markcite{eem}Elmegreen et al.
1992).  

We find that {\it internal perturbations}
by ``massive disturbers'' orbiting
{\it within} the disk seem to do
the best job at replicating the crooked FUV arms observed
in M101.  Simulations explicitly including massive disturbers 
in the disk (\markcite{byrd}Byrd 1983; \markcite{bsm}Byrd,
Smith, \& Miller 1984), as
well as simulations which implicitly include them
through large gas/star mass ratios and  
strong spiral forcing (leading to massive agglomerations) 
(\markcite{r92}Roberts 1992; \markcite{r93}Roberts 1993)
yield spiral arm patterns with linear
arm segments and with opening 
angles of approximately 120$^{\circ}$.

\subsection {Summation}

Based on the above comparisons, 
we speculate that M101 has been subject to tidal
perturbations from one or more companion galaxies in the past 10$^8$--10$^9$
years.  The interaction(s) have reshaped
the large-scale spiral structure in the
disk of M101, including the generation of a tidal tail to the south and
a massive condensation (NGC 5471) at the end of the tail to the east that is 
currently undergoing a major starburst.  Lower levels of ongoing star formation 
are evident along the length of the tidal tail.
Further
perturbations by NGC 5471 and other ``massive disturbers'' orbiting in the disk
of M101 have induced
a disk-wide pattern of crooked arms with linear arm segments.  

The prominence
of the crooked-arm pattern at FUV wavelengths indicates that it traces the most
recent epoch of star formation and hence is a transient phenomenon.
Consideration of the shear flow in the disk of M101 as approximated by 

$$\Delta v \ = \ \Delta R \ R \ d\Omega / dR \ \approx \ 
-v_{rot} \ \Delta R / R,$$

\noindent
where $d\Omega / dR$ is the differential rotation in units of km s$^{-1}$ 
kpc$^{-2}$ 
(cf. \markcite{w90}Waller 1990; \markcite{bt}Binney \& Tremaine 1987), 
indicates that significant evolution of the crooked-arm 
pattern will occur in a time less than 20 Myr --- thus explaining the 
exceptional clarity of the FUV arms compared to the more diffuse spiral 
structures evident at 
longer wavelengths.  

The proposed 
tidal scenario for explaining the unusual morphological and kinematic features
in M101 can be tested via numerical simulations of the M101 system of galaxies 
(cf. \markcite{cow}Combes et al. 1996)
and by evaluating the properties of other giant spiral galaxies
that display similar features (e.g. NGC 1232, NGC 2805, \& NGC 4303).

\acknowledgements 
We thank Jeff Kenney for access to his CO map in advance of publication.
Gene Byrd, Francoise Combes, Kevin Olson, and Bill Roberts
generously provided insights on the dynamics of interacting disk galaxies and
assistance in diagnosing M101's unusual properties.  UIT research is funded
through the Spacelab Office at NASA Headquarters under Project number 440-51.
We are deeply grateful to the crew of STS-67 and to the many people who helped
make the Spacelab/{\it Astro-2} mission such a great success.

\begin{deluxetable}{cccccccccc}
\tablewidth{40pc}
\footnotesize
\tablecaption{Linear Arm Segments}

\tablehead{\colhead{No.} & \colhead{$<PA>$} & \colhead{$<R>$} & 
\colhead{$<Pitch>$}
& \colhead{$\Delta Pitch$} & \colhead{$Length$} & \colhead{$<I(FUV)>$} & 
\colhead{$Contrast$} \\[.2ex]
\colhead{} & \colhead{(deg)} & \colhead{(kpc)} & \colhead{(deg)} & 
\colhead{(deg)} 
& \colhead{(kpc)} & \colhead{} 
& \colhead{}  \\[.2ex]
\colhead{(1)} & \colhead{(2)} & \colhead{(3)} & \colhead{(4)} 
& \colhead{(5)} & \colhead{(6)} & \colhead{(7)} & \colhead{(8)}} 
\startdata

1&  312&     22.3&     2&      -22&   8.6&   323&   4.7 \nl

2&  331&     21.0&     24&     -18&   6.9&   342&   5.4 \nl

3&  353&     7.8&      7&      -80&  10.2&   594&   3.1 \nl

4&  195&     6.5&      30&     -51&   6.6&   534&   1.7 \nl

5&  168&     7.5&      16&     106&  12.5&   365&   1.3 \nl

6&  156&     9.3&      21&      27&   4.6&   943&   3.5 \nl

7&   50&     5.8&      22&     -43&   4.7&   498&   1.9 \nl

8&   42&     8.8&      27&     -47&   8.9&   918&   2.7 \nl

9&   31&     15.6&     35&     -42&  13.6&   100&   2.6 \nl

10& 120&     11.0&     10&      45&   8.5&   1030&   5.0 \nl

11&  40&     21.8&     23&     -57&  22.6&   228&   5.2 \nl
   
12&  81&     14.6&     34&      34&  10.4&   965&   6.7 \nl

\enddata

\tablecomments{All quantities are measured in the plane of M101's disk, 
assuming an inclination of $i = 18^{\circ}$ and major axis position angle of 
$PA = 39^{\circ}$ (Bosma et al. 1981).  
(1) --- Linear arm segment number (see {\bf Figure 2a}).
(2) --- Mean position angle (in degrees)
of arm segment, where the major axis has $PA = 39^{\circ}$, as observed on the 
sky.
(3) --- Mean galactocentric radius (in kpc) of the arm segment, 
where the distance to M101 is
assumed to be 7.4 Mpc.  For reference, the Holmberg
radius is 30.1 kpc, and the radius of NGC 5471 is 25.1 kpc.
(4) --- Mean pitch angle (in degrees) of
arm segment with respect to a
circle corresponding to the mean radius.
(5) --- Range of pitch angles (in degrees) along arm segment as measured from 
west to east.
(6) --- Length of arm segment (in kpc).
(7) --- Average FUV 
intensity (surface brightness) in units of 
10$^{-20}$ erg cm$^{-2}$ s$^{-1}$ arcsec$^{-2}$ 
\AA$^{-1}$, based on strip scans of 45$''$ (1.6 kpc) width on and off each arm.
(8) --- Arm--interarm contrast from the ratio of median intensities
on and off each arm segment.}
\end{deluxetable}

\figcaption{(Plate xxx) Far-ultraviolet ($\lambda_{\circ}$1521,
$\Delta\lambda$354) image of M101 as obtained by
the Ultraviolet Imaging Telescope during the Spacelab/{\it Astro-2} mission of
March 1995.  The image has been corrected for residual field distortion.
North is up
and East is to the left.
The
astrometry is based on comparison of point sources with corresponding sources
in the Guide Star Catalog.
After rotation and centering of the target,
the unmasked field of view has a diameter of 32.3 arcmin (69.5 kpc).}

\figcaption{(Plate xxx) {\bf (a)} --- Far-ultraviolet image of M101 after
spatial filtering to highlight the 
intermediate-scale structure.  FUV arm segments 
described in the text are delineated and labeled (see {\bf Table 1}).
{\bf (b.)}
High-contrast presentation of the FUV emission
in M101.  The image has been Gaussian smoothed to a resolution of 10$''$
(FWHM).  This was done to highlight the faint southern arm (``Arm'')
and its looping
connection with NGC 5471 to the east.  The FUV knots (``T1, T2, T3, T4'')
that help delineate the
``curly tail'' feature have H$\alpha$ counterparts as described in the text.}

\figcaption
{(Plate xxx) 
Superposition of contoured CO emission (Kenney
et al. 1996) on a grey-scale image of the FUV emission in the central $4'
\times 4'$ (8.6 $\times$ 8.6 kpc) of M101.}

\end{document}